\documentclass{article}
\usepackage{spconf,amsmath,graphicx}

\usepackage{multirow}
\usepackage{graphicx}
\usepackage{amssymb, amsmath, bm, mathtools,array}
\usepackage{textcomp}
\usepackage{hyperref}
\usepackage{verbatim,lipsum}
\usepackage{booktabs}
\usepackage{tcolorbox}
\usepackage{lscape}
\usepackage{subcaption}
\usepackage{multirow}
\usepackage{colortbl}
\usepackage{color}
\usepackage{multibib}

\RequirePackage{color}\definecolor{BLUE}{rgb}{0,0,0}

\newcommand{\setTLAshort}{{\texttt{LA19trn}}}

\newcommand{\setTLAVOC}{{\texttt{voc.LA}}}
\newcommand{\setTVOXVOC}{{\texttt{voc.VoxCel}}}

\newcommand{\setELA}{{\texttt{LA19eval}}}
\newcommand{\setELAII}{{\texttt{LA21eval}}}
\newcommand{\setEDF}{{\texttt{DF21eval}}}
\newcommand{\setELATRIM}{{\texttt{LA19etrim}}}
\newcommand{\setELAHID}{{\texttt{LA21hid}}}
\newcommand{\setEDFHID}{{\texttt{DF21hid}}}
\newcommand{\setEWFE}{{\texttt{WaveFake}}}
\newcommand{\setEDFE}{{\texttt{InWild}}}

\newcommand{\setGAll}{{\texttt{Pooled}}}

\newcommand{\sslxlsr}{{\textsf{xlsr}}}
\newcommand{\ssllv}{{\textsf{w2v}}}
\newcommand{\sslvox}{{\textsf{v.vox}}}

\newcommand{\selfcircle}[1]{\textsf{#1}}

\newcites{app}{Appendix References}

\title{Can large-scale vocoded spoofed data improve speech spoofing countermeasure with a self-supervised front end?}
\name{Xin Wang\thanks{This work was supported by JST CREST Grants JPMJCR18A6 and JPMJCR20D3, JST PRESTO Grant number JPMJPR23P9, MEXT KAKENHI Grants 21K17775 and 21H04906.}, Junichi Yamagishi}
\address{National Institute of Informatics, Japan}

\begin{document}
\ninept
\onecolumn
{\noindent\Large \textbf{IEEE Copyright Notice}}

${}$

{\noindent\large \copyright 2024 IEEE. 
Personal use of this material is permitted. Permission from IEEE must be obtained for all other uses, in any current or future media, including reprinting/republishing this material for advertising or promotional purposes, creating new collective works, for resale or redistribution to servers or lists, or reuse of any copyrighted component of this work in other works.

${}$

\noindent
This work is accepted by the IEEE International Conference on Acoustics, Speech and Signal Processing.

${}$

\noindent
}
\twocolumn

%
\maketitle
\begin{abstract}
A speech spoofing countermeasure (CM) that discriminates between unseen spoofed and bona fide data requires diverse training data. 
While many datasets use spoofed data generated by speech synthesis systems, it was recently found that data vocoded by neural vocoders were also effective as the spoofed training data.  
Since many neural vocoders are fast in building and generation, this study used multiple neural vocoders and created more than 9,000 hours of vocoded data on the basis of the VoxCeleb2 corpus. 
This study investigates how this large-scale vocoded data can improve spoofing countermeasures that use data-hungry self-supervised learning (SSL) models.
Experiments demonstrated that the overall CM performance on multiple test sets improved when using features extracted by an SSL model continually trained on the vocoded data. Further improvement was observed when using a new SSL distilled from the two SSLs before and after the continual training. The CM with the distilled SSL outperformed the previous best model on challenging unseen test sets, including the ASVspoof 2019 logical access, WaveFake, and In-the-Wild.
\end{abstract}
\begin{keywords}
anti-spoofing, presentation attack detection, countermeasure, logical access, neural vocoder
\end{keywords}

\section{Introduction}
\label{sec:intro}

The detection of spoofed speech generated by text-to-speech (TTS) and voice conversion (VC) systems is usually formulated as a binary classification task \cite{Wu2015}. A detector, referred to as a spoofing countermeasure (CM), requires a significant amount of training data containing diverse human (bona fide) and synthesized (spoofed) speech waveforms. However, preparing various spoofed training data is costly. For example, it took a few months to develop the TTS and VC systems that generated the training set of the ASVspoof 2019 logical access (LA) database through trial and error \cite{asvspoof2019_database}. 

Instead of TTS or VC,  one study used neural network (NN)-based vocoders to create spoofed data through vocoding or copy-synthesis \cite{Wang2023a,Sun2023CVPR,sanchez2014cross,sizov2015joint}. It extracts acoustic features from the bona fide data and converts them into vocoded spoofed waveforms using the vocoders. Since most TTS and VC systems end with a vocoder or vocoder-like module, the vocoded data is expected to capture the vocoders' artifacts that affect TTS and VC systems. 
The vocoders are straightforward to train, and those in \cite{Wang2023a} took less than one week to build.
The CMs trained using bona fide and vocoded spoofed data performed well in detecting unseen TTS and VC. 

Vocoding makes producing a massive amount of data feasible, but the previous work only created a small set of 10 hours \cite{Wang2023a}. This study pushed the limits and created more than 9,000 hours of data vocoded from the VoxCeleb2 database \cite{chung18b_interspeech}. With this large-scale vocoded spoofed data, this study investigated how we can improve the CM performance through experiments.

This study focuses on CMs based on self-supervised learning (SSL) speech models because they performed the best on vocoded data \cite{Wang2023a}. Two use cases for the large-scale vocoded data are investigated: 1) the vocoded spoofed data is used to train the SSL in a self-supervised manner, after which the SSL is used in feature extraction for the CM; 2) the vocoded spoofed and bona fide data are used to fine-tune the entire CM in a supervised manner. In the first use case, because training the SSL model from scratch is infeasible on our computation platform, this study compared three configurations: 
1.1) using the features extracted from the SSL initialized by a pre-trained one and continually trained on the vocoded data for a few epochs, 
1.2) using the feature differences between the continually trained and pre-trained SSLs, 
1.3) using the features extracted from a new SSL distilled from the two SSLs.

As the experiment results demonstrated, the continually trained SSL improved the performance on multiple test sets. 
Using the feature differences between the two SSLs further improved the results, while using the distilled SSL reduced the model size without sacrificing the overall performance. 
In contrast, using the large-scale vocoded data to fine-tune the CM in a supervised manner degraded the performance on some of the test sets significantly. 

We understand that the cost of training the SSL models may offset the improvement in the CM, 
but the findings in this study may be helpful to other practitioners who want to avoid pitfalls.
We have also released the SSL models trained on the vocoded VoxCeleb2 data and the trained CMs\footnote{The link to the code repository and the paper appendix are available on Arxiv page \url{https://arxiv.org/abs/2309.06014}.}.
Related works \cite{mun2023towards, zeng2023improving} used a similar approach and created a large amount of vocoded data, but they focused on either spoofing-aware speaker verification or CM without using SSL-based models. As far as we know, this study is the first to combine large-scale vocoded data and SSL-based CMs.

\section{Methods}
\label{sec:method}

\subsection{Creating vocoded spoofed data}
\label{sec:vocoder}

This study followed \cite{Wang2023a} to create vocoded data.
The neural vocoders include a general adversarial network-based vocoder called HiFiGAN \cite{NEURIPS2020_c5d73680}, a flow-based vocoder called WaveGlow \cite{prenger2018waveglow}, a hybrid vocoder called the harmonic-plus-noise neural source-filter model (Hn-NSF) \cite{wangNSFall} that combines deep learning and digital signal processing techniques, and a fusion of Hn-NSF and HiFiGAN. The vocoders were selected because they can generate speech waveforms faster than real-time speed.  Other neural vocoders that generate waveforms auto-regressively are not considered due to the unbearable time cost to produce a large amount of vocoded data.

Pre-trained vocoders released by \cite{Wang2023a} were downloaded and fine-tuned on the VoxCeleb2 development set using the provided recipes. With the fine-tuned neural vocoders, we extracted acoustic features from the data in the VoxCeleb2 development set and drove the vocoders to synthesize the vocoded waveforms. The acoustic features include an 80-dimensional Mel spectrogram and, optionally, 1-dimensional fundamental frequency.  We collected 2.3k hours of vocoded data from each of the four vocoders, resulting in a total of 9.2k hours of vocoded data.

\subsection{Using large-scale vocoded data on SSL-based CMs}
This study applies the large-scale vocoded data to SSL-based CMs because they outperformed other CMs when using the vocoded data \cite{Wang2023a}. In such a CM, an SSL-based model first extracts a sequence of hidden feature vectors from an input waveform. An NN-based back end then converts the hidden features into a score that indicates how likely the input is bona fide.  This section describes how the large-scale vocoded data is used for the SSL-based CM.

\begin{figure}[t!]
\begin{center}
\includegraphics[trim=0 10 0 0, width=\columnwidth]{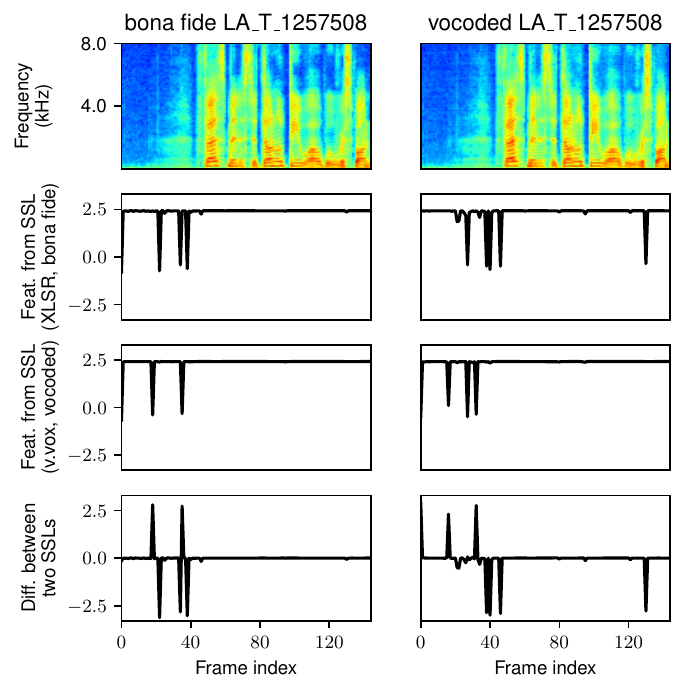}
\caption{Spectrograms of bona fide and vocoded utterances (top row), output features from Wav2Vec 2.0 pre-trained on bona fide data (2nd row) and that continually trained on vocoded VoxCeleb2 data (3rd row), and their differences (bottom row). The two SSLs correspond to \sslxlsr\ and \sslvox\ in Section~\ref{sec:exp1}, respectively. 
The plotted feature dimension corresponds to the one with the largest variance in the differential features between the two SSLs. 
}
\label{fig_twossl}
\vspace{-5mm}
\end{center}
\end{figure}

\begin{figure}[t!]
\begin{center}
\includegraphics[trim=0 560 0 60, width=\columnwidth]{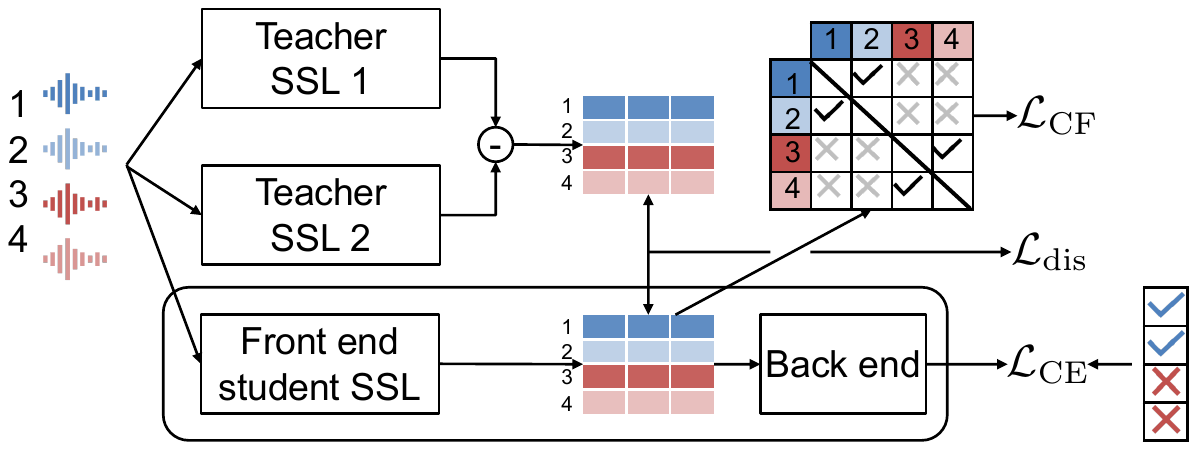}
\caption{Training CM while distilling SSL. 
Data 1, 2, 3, and 4 are bona fide, bona fide after data augmentation, vocoded spoofed, and its augmented version, respectively.   
$\checkmark$ and $\times$ for $\mathcal{L}_{\text{CF}}$ denote the same and different views for supervised contrastive learning, respectively.}
\label{fig_cr}
\vspace{-5mm}
\end{center}
\end{figure}

\subsubsection{Upstream continual self-supervised training of SSL}
A well-performed SSL is typically trained using a self-supervised training criterion \cite{Mohamed2022} on an extensive amount of bona fide speech data and computational resources (e.g., 128 Tesla V100 GPUs \cite{NEURIPS2020_92d1e1eb}). This process is referred to as the upstream training of SSL \cite{Mohamed2022}.

Since the SSL is to be used for anti-spoofing in this study, it is reasonable to perform the upstream training using spoofed data. This enables the SSL to better detect artifacts in the spoofed data.  Unfortunately, because it is infeasible to train the SSL from scratch on our GPU platform, we continually train the SSL. This is achieved by loading an SSL model pre-trained on bona fide data and continually training it on the vocoded data using the self-supervised training criterion.

Figure~\ref{fig_twossl} shows the extracted features from bona fide and vocoded waveforms using a Wav2Vec2.0 model pre-trained on bona fide data \cite{NEURIPS2020_92d1e1eb} and its variant continually trained on the vocoded VoxCeleb2 data.
The features are generated by the layer-normalization layer after the last transformer block. 
It can be observed that the continually trained SSL produces a spike around the 120th frame of the vocoded waveform (3rd row, right column), which corresponds to a consonant sound. The two SSL models may generate slightly different and complementary features.

\subsubsection{Downstream fine-tuning of SSL-based CM}
After upstream training, it is recommended to fine-tune the SSL with the CM using both bona fide and spoofed (or vocoded spoofed) data in a supervised manner \cite{tak2022automatic, wang2021investigating,martin2022vicomtech}.  
While existing work \cite{Wang2023a} used the ASVspoof 2019 LA training set \cite{asvspoof2019_database} or its vocoded version, this study investigates the potential benefits of fine-tuning the CM using VoxCeleb2 and its vocoded data.
Following the methodology of \cite{Wang2023a}, this study conducts downstream fine-tuning using the standard cross-entropy loss ($\mathcal{L}_{\text{CE}}$) and a contrastive feature loss ($\mathcal{L}_{\text{CF}}$). 
The latter is computed over the frame-level features extracted by the SSL model from a bona fide utterance, its vocoded version, and their data-augmented variants, as shown in Figure~\ref{fig_cr}.  

\subsubsection{Distilling a single SSL during downstream fine-tuning}
Figure~\ref{fig_twossl} shows that the pre-trained and continually trained SSLs yield similar feature trajectories.
By calculating their difference, we derive a feature trajectory that centers around zero and contains the information from both SSLs (last row in Figure~\ref{fig_twossl}). We hypothesize that the differential features are beneficial to the CM.

One approach is to integrate the two SSLs into the CM and compute the differential features on-the-fly, but this doubles the CM's model size. We propose distilling the two SSLs into a single SSL, as illustrated in Figure~\ref{fig_cr}. 
Let the outputs of two teacher SSLs be ${\boldsymbol{x}}_{1:N}=(\boldsymbol{x}_1, \cdots, \boldsymbol{x}_N)$ and $\tilde{\boldsymbol{x}}_{1:N}$, where $N$ denotes the number of feature frames. Let ${\boldsymbol{z}}_{1:N}=(\boldsymbol{z}_1, \cdots, \boldsymbol{z}_N)$ denote the output of the student SSL in the CM. A distillation loss can be defined as
\begin{equation}
\mathcal{L}_{\text{dis}}({\boldsymbol{x}}_{1:N}, \tilde{\boldsymbol{x}}_{1:N}, {\boldsymbol{z}}_{1:N}) = {\frac{1}{N}}\sum_{i=1}^{N}\Big|\Big| \boldsymbol{z}_i - |\boldsymbol{x}_i - \tilde{\boldsymbol{x}}_i | \Big|\Big|_1.
\end{equation}
The student SSL is fine-tuned with the rest of the CM using the training loss $\mathcal{L}_{\text{CE}} + \lambda_{\text{CF}}\mathcal{L}_{\text{CF}}+\lambda_{\text{dis}}\mathcal{L}_{\text{dis}}$.

This study uses the configuration described in \cite{Wang2023a} and sets $\lambda_{\text{CF}}=1$. 
It then sets $\lambda_{\text{dis}}=100$ so that the amplitude of  $\lambda_{\text{dis}}\mathcal{L}_{\text{dis}}$ is comparable to that of  $\lambda_{\text{CF}}\mathcal{L}_{\text{CF}}$.  Using a larger or smaller $\mathcal{L}_{\text{dis}}$ degraded the overall performance. 
Another useful configuration is initializing the student SSL using one of the teacher SSLs. 
Random initialization significantly degrades the CM's performance. 
Results of the ablation study can be found in the appendix on Arxiv.

\begin{table*}[t!]
\caption{Test set EERs (\%). Darker cell color indicates a higher EER value. 
\selfcircle{B1-b} and \selfcircle{B1-c} are the same as \selfcircle{B1} except that they were fine-tuned on \setTLAshort\ and \setTVOXVOC, respectively. Similarly, \selfcircle{P3-b} and \selfcircle{P3-c} are variants of \selfcircle{P3}.
}
\vspace{-6mm}
\begin{center}
\footnotesize
\setlength{\tabcolsep}{8pt}
\resizebox{\textwidth}{!}
{
\begin{tabular}{rr||rrr|rrr||rr|rr}
\toprule
      &      & \multicolumn{6}{c||}{ Section~\ref{sec:exp1} }   &   \multicolumn{4}{c}{Section~\ref{sec:exp2} }  \\
      \midrule
 \multirow{3}{*}{\rotatebox{90}{CM}}       &          ID          &  \selfcircle{B1}   & \selfcircle{B2}          & \selfcircle{B3}  &  \selfcircle{P1} & \selfcircle{P2} & \selfcircle{P3}  & \selfcircle{B1-b}   & \selfcircle{P3-b}          & \selfcircle{B1-c}  &  \selfcircle{P3-c} \\ 
\cmidrule{2-12}
& {Front end SSL(s)} & 
\textcolor{white}{-\ssllv}\sslxlsr & \sslxlsr, \ssllv   & \sslxlsr, \ssllv & \textcolor{white}{-\sslxlsr}\sslvox & \sslxlsr, \sslvox & \sslxlsr, \sslvox  & \textcolor{white}{-\ssllv}\sslxlsr & \sslxlsr, \sslvox  & \textcolor{white}{-\ssllv}\sslxlsr & \sslxlsr, \sslvox    \\
 &  
SSL distilling & - &  $\times$  & $\checkmark$ &  - & $\times$  & $\checkmark$  & -  & $\checkmark$ & - & $\checkmark$ \\
\midrule
\multicolumn{2}{r||}{Data for fine-tune CM} & 
  \multicolumn{3}{c|}{ \setTLAVOC} &  \multicolumn{3}{c||}{ \setTLAVOC}   &  \multicolumn{2}{c|}{ \setTLAshort}  &   \multicolumn{2}{c}{ \setTVOXVOC}  \\
\midrule
\midrule
\multirow{11}{*}{\rotatebox{90}{Test sets}}   &   \setELA   & \cellcolor[rgb]{0.96, 0.96, 0.96} 3.45 & \cellcolor[rgb]{0.98, 0.98, 0.98} 1.97 & \cellcolor[rgb]{0.99, 0.99, 0.99} 1.26 & \cellcolor[rgb]{0.98, 0.98, 0.98} 2.09 & \cellcolor[rgb]{0.98, 0.98, 0.98} 2.01 & \cellcolor[rgb]{0.98, 0.98, 0.98} 1.91 & \cellcolor[rgb]{1.00, 1.00, 1.00} 0.22 & \cellcolor[rgb]{1.00, 1.00, 1.00} 0.13 & \cellcolor[rgb]{0.96, 0.96, 0.96} 3.59 & \cellcolor[rgb]{0.96, 0.96, 0.96} 3.71\\ 
  &  \setELAII  & \cellcolor[rgb]{0.66, 0.66, 0.66} 17.59 & \cellcolor[rgb]{0.76, 0.76, 0.76} 13.94 & \cellcolor[rgb]{0.59, 0.59, 0.59} 21.09 & \cellcolor[rgb]{0.68, 0.68, 0.68} 16.88 & \cellcolor[rgb]{0.74, 0.74, 0.74} 14.94 & \cellcolor[rgb]{0.72, 0.72, 0.72} 15.92 & \cellcolor[rgb]{0.97, 0.97, 0.97} 2.69 & \cellcolor[rgb]{0.96, 0.96, 0.96} 3.29 & \cellcolor[rgb]{0.73, 0.73, 0.73} 15.22 & \cellcolor[rgb]{0.80, 0.80, 0.80} 12.37\\ 
  &   \setEDF   & \cellcolor[rgb]{0.92, 0.92, 0.92} 6.53 & \cellcolor[rgb]{0.95, 0.95, 0.95} 4.04 & \cellcolor[rgb]{0.75, 0.75, 0.75} 14.72 & \cellcolor[rgb]{0.95, 0.95, 0.95} 4.34 & \cellcolor[rgb]{0.94, 0.94, 0.94} 5.28 & \cellcolor[rgb]{0.93, 0.93, 0.93} 5.67 & \cellcolor[rgb]{0.95, 0.95, 0.95} 4.27 & \cellcolor[rgb]{0.96, 0.96, 0.96} 3.45 & \cellcolor[rgb]{0.92, 0.92, 0.92} 5.99 & \cellcolor[rgb]{0.96, 0.96, 0.96} 3.31\\ 
        \cmidrule{2-12}
  & \setELATRIM & \cellcolor[rgb]{0.97, 0.97, 0.97} 2.69 & \cellcolor[rgb]{0.97, 0.97, 0.97} 2.80 & \cellcolor[rgb]{0.96, 0.96, 0.96} 3.74 & \cellcolor[rgb]{0.96, 0.96, 0.96} 3.33 & \cellcolor[rgb]{0.97, 0.97, 0.97} 2.79 & \cellcolor[rgb]{0.96, 0.96, 0.96} 3.28 & \cellcolor[rgb]{0.90, 0.90, 0.90} 7.37 & \cellcolor[rgb]{0.90, 0.90, 0.90} 7.37 & \cellcolor[rgb]{0.97, 0.97, 0.97} 2.74 & \cellcolor[rgb]{0.96, 0.96, 0.96} 3.63\\ 
  & \setELAHID  & \cellcolor[rgb]{0.76, 0.76, 0.76} 13.93 & \cellcolor[rgb]{0.76, 0.76, 0.76} 14.05 & \cellcolor[rgb]{0.59, 0.59, 0.59} 20.03 & \cellcolor[rgb]{0.71, 0.71, 0.71} 16.02 & \cellcolor[rgb]{0.76, 0.76, 0.76} 13.95 & \cellcolor[rgb]{0.74, 0.74, 0.74} 14.97 & \cellcolor[rgb]{0.72, 0.72, 0.72} 15.56 & \cellcolor[rgb]{0.59, 0.59, 0.59} 24.23 & \cellcolor[rgb]{0.85, 0.85, 0.85} 10.14 & \cellcolor[rgb]{0.86, 0.86, 0.86} 9.53\\ 
  & \setEDFHID  & \cellcolor[rgb]{0.87, 0.87, 0.87} 8.89 & \cellcolor[rgb]{0.87, 0.87, 0.87} 9.10 & \cellcolor[rgb]{0.73, 0.73, 0.73} 15.27 & \cellcolor[rgb]{0.89, 0.89, 0.89} 7.71 & \cellcolor[rgb]{0.88, 0.88, 0.88} 8.40 & \cellcolor[rgb]{0.87, 0.87, 0.87} 8.84 & \cellcolor[rgb]{0.87, 0.87, 0.87} 9.16 & \cellcolor[rgb]{0.76, 0.76, 0.76} 13.95 & \cellcolor[rgb]{0.87, 0.87, 0.87} 9.03 & \cellcolor[rgb]{0.89, 0.89, 0.89} 7.77\\ 
  &  \setEWFE   & \cellcolor[rgb]{0.90, 0.90, 0.90} 7.33 & \cellcolor[rgb]{0.98, 0.98, 0.98} 1.48 & \cellcolor[rgb]{0.93, 0.93, 0.93} 5.88 & \cellcolor[rgb]{0.98, 0.98, 0.98} 1.94 & \cellcolor[rgb]{0.99, 0.99, 0.99} 0.89 & \cellcolor[rgb]{0.99, 0.99, 0.99} 1.30 & \cellcolor[rgb]{0.59, 0.59, 0.59} 23.75 & \cellcolor[rgb]{0.73, 0.73, 0.73} 15.44 & \cellcolor[rgb]{0.78, 0.78, 0.78} 13.41 & \cellcolor[rgb]{0.59, 0.59, 0.59} 24.17\\ 
  &  \setEDFE   & \cellcolor[rgb]{0.91, 0.91, 0.91} 6.78 & \cellcolor[rgb]{0.95, 0.95, 0.95} 4.25 & \cellcolor[rgb]{0.78, 0.78, 0.78} 13.20 & \cellcolor[rgb]{0.93, 0.93, 0.93} 5.84 & \cellcolor[rgb]{0.95, 0.95, 0.95} 4.07 & \cellcolor[rgb]{0.92, 0.92, 0.92} 6.10 & \cellcolor[rgb]{0.77, 0.77, 0.77} 13.52 & \cellcolor[rgb]{0.80, 0.80, 0.80} 12.32 & \cellcolor[rgb]{0.91, 0.91, 0.91} 6.90 & \cellcolor[rgb]{0.91, 0.91, 0.91} 7.00\\ 
        \cmidrule{2-12}
  &  \setGAll   & \cellcolor[rgb]{0.83, 0.83, 0.83} 11.13 & \cellcolor[rgb]{0.79, 0.79, 0.79} 12.95 & \cellcolor[rgb]{0.76, 0.76, 0.76} 14.06 & \cellcolor[rgb]{0.84, 0.84, 0.84} 10.54 & \cellcolor[rgb]{0.87, 0.87, 0.87} 9.07 & \cellcolor[rgb]{0.85, 0.85, 0.85} 9.98 & \cellcolor[rgb]{0.79, 0.79, 0.79} 12.76 & \cellcolor[rgb]{0.80, 0.80, 0.80} 12.50 & \cellcolor[rgb]{0.83, 0.83, 0.83} 10.92 & \cellcolor[rgb]{0.80, 0.80, 0.80} 12.26\\ \bottomrule
\end{tabular}
}
\label{tab_eer}
\vspace{-6mm}
\end{center}
\end{table*}

\section{Experiments}
\label{sec:exp}

\subsection{Datasets}
\label{sec:database}


\textbf{Training sets:} As described in Section~\ref{sec:vocoder}, this study used the VoxCeleb2 development set to create the large-scale vocoded data. This dataset contains 5,994 speakers' speech data with a total duration of 2,360 hours. 
The four neural vocoders mentioned in Section~\ref{sec:vocoder}  were fine-tuned on this dataset for ten epochs at a sampling rate of 16 kHz using the recipe outlined in \cite{Wang2023a}. Each vocoder was used to vocode the VoxCeleb2 development set.  The total duration of the vocoded data reached 9,440 hours.
This vocoded data is used in the upstream continual training of the SSL or downstream CM fine-tuning.

For downstream CM fine-tuning, this study also compared the ASVspoof 2019 LA training set (\setTLAshort) and a small vocoded dataset created from it, \setTLAVOC\footnote{It is the training set \texttt{Voc.v4} in \cite{Wang2023a}.}. The same four types of neural vocoders were used to create \texttt{Voc.v4}, but they were fine-tuned on ASVspoof bona fide data. Both datasets cover data from 20 speakers; their durations are around 24 and 10 hours, respectively.

\textbf{Test sets:} This study used multiple test sets to measure the CM performance in both seen and unseen domains. They included the official evaluation sets of ASVspoof 2019 LA  (\setELA), ASVspoof 2021 LA (\setELAII), and the 2021 DF (\setEDF) tasks. 
Additionally, this study used the hidden subsets of the ASVspoof 2021 LA ({\setELAHID}) and DF ({\setEDFHID}) datasets \cite{LiuASVspoof2021} and a variant of the ASVspoof 2019 LA evaluation set ({\setELATRIM}) provided by \cite{Wang2023a}, in which non-speech segments at the two ends of each utterance are removed. 
These hidden subsets are expected to measure the CM performance in a more challenging scenario, where the CM cannot exploit discriminative features from non-speech segments, for example,  the unbalanced distribution of the length of non-speech segments \cite{muller21_asvspoof} or energy levels in different frequency bands \cite{zhang21da_interspeech}.

Two out-of-domain test sets were WaveFake (\setEWFE) \cite{frank2021wavefake} and the In-the-Wild datasets (\setEDFE) \cite{muller2022does}, both of which are from domains different from ASVspoof datasets. In particular, {\setEDFE} contains bona fide and spoofed data from over 50 English-speaking celebrities, collected from the Internet. A generalizable CM should perform well on these test sets in a zero-shot manner.

\subsection{Training recipe and evaluation metric}
The experimental CMs consist of an SSL-based front end and a neural-network-based back end. 
All the CMs used the same back-end architecture, including a global average pooling layer, three fully-connected layers with LeakyReLU activation functions, and a linear layer that outputs a score.  
They all used RawBoost with its best configuration in \cite{Tak2021} for data augmentation. The CMs differ in terms of the front-end and training data configurations.

During downstream fine-tuning, all the CMs used the Adam optimizer ($\beta_1=0.9, \beta_2=0.999, \epsilon=10^{-8}$) \cite{kingma2014adam}. The learning rate was initialized to $5\times10^{-6}$ and reduced by a factor of $0.1$ every ten epochs. 
The training process was halted when the validation loss on the ASVspoof 2019 LA development set failed to improve within ten epochs. 
Each training utterance was randomly truncated with a maximum duration of 4s to fit the GPU memory. During inference, the entire test utterance was processed without truncation.

CM performance was measured using the equal error rate (EER) on both individual and pooled test sets.  
Each CM was trained and evaluated for three rounds, and each round used a different random seed for network initialization except for the SSL-based front end. The average EER across the three rounds is reported.

\subsection{Experiment 1: large-scale vocoded data for SSL front end}
\label{sec:exp1}

This experiment investigated the usage of large-scale vocoded data on the SSL-based front ends and compared the following CMs: 
\begin{itemize}
\item \selfcircle{B1}'s front end is a Wav2Vec-2.0-based model called XLSR-53 and pre-trained on multilingual speech data \cite{conneau21_interspeech}. This SSL is denoted as  \sslxlsr\ in this study. 
\item \selfcircle{P1} was configured in the same manner as \selfcircle{B1}, but the SSL was continually trained from \sslxlsr\ on the vocoded VoxCeleb2 data. This new SSL model is referred to as \sslvox. 
\item \selfcircle{B2} uses \sslxlsr\ and another Wav2Vec-2.0 model \ssllv\ pre-trained on four English databases\footnote{Fairseq (https://github.com/facebookresearch/fairseq) Wav2Vec 2.0 Large (LV-60+CV+SWBD+FSH) model, no ASR finetuning.}. \selfcircle{B2} uses the differences between \sslxlsr\ and \ssllv's output as the back-end input.
\item \selfcircle{P2} was configured in the same manner as \selfcircle{B2} except that the two SSLs are \sslxlsr\ and \sslvox.
\item \selfcircle{B3} uses a student SSL initialized by  \sslxlsr\ and distilled from \sslxlsr\ and \ssllv. 
\item \selfcircle{P3} was the same as \selfcircle{B3} but distilled from \sslxlsr\ and \sslvox.
\end{itemize}
All the CMs were fine-tuned on \setTLAVOC. 
Note that \selfcircle{B1} was configured and trained in the same manner as the best CM in \cite{Wang2023a}. 
The SSL's continuous training was conducted using the Fairseq toolkit \cite{ott2019fairseq}  for three epochs, with a peak learning rate of 1e-04. This training stage took six days using four Nvidia A100 GPU cards.

The results are listed in the left part of Table~\ref{tab_eer}.  The comparison between \selfcircle{B1} and \selfcircle{P1} shows that using the SSL  \sslvox\ continually trained on the vocoded data helps to reduce the pooled EER from 11.13\% to 10.54\%. EERs on each test set improved, except for \setELAHID\ and \setELATRIM. This suggests that using the continually trained SSL is beneficial to the CM.

The comparison between \selfcircle{P1} and \selfcircle{P2} shows that using the continually trained and pre-trained SSLs together further reduced the pooled EER to 9.07\%. Although the EERs on \setEDF\ and \setEDFHID\ became slightly higher, the lower pooled EER indicates it is effective to use the differential features between the two SSLs.   
Readers may argue that the improvement from \selfcircle{P1} to \selfcircle{P2} is due to the increased model size. However, \selfcircle{B2}, which has the same model size as \selfcircle{P2}, obtained a higher pooled EER of 12.95\%. 
It was observed that the outputs from \sslxlsr\ and \ssllv\ used by \selfcircle{B2} were quite different, and their difference was not as sparse as that from \sslxlsr\ and \sslvox\ (see appendix on Arxiv).
Hence, pairing two SSLs before and after the continual training (e.g., \sslxlsr\ and \sslvox)  might be better than merging SSLs trained on different bona fide data (e.g., \sslxlsr\ and \ssllv). 

Note that \selfcircle{B2}'s EERs on individual test sets were lower than those of \selfcircle{B1} in many cases, but \selfcircle{B2}'s pooled EER is higher. This is because scores produced by \selfcircle{B2} on different test sets are poorly aligned. A single classification threshold cannot effectively separate bona fide and spoofed samples across the test sets.

The comparison between \selfcircle{P2} and \selfcircle{P3} demonstrates that the differences between \sslvox\ and \sslxlsr\ can be distilled into a single SSL to a certain degree. We found that the feature extracted by the student SSL was close to the feature plotted at the bottom of Figure~\ref{fig_twossl}. Although \selfcircle{P3}'s pooled EER was slightly higher than that of \selfcircle{P2}, both outperformed \selfcircle{B1}, which is the best CM in \cite{Wang2023a}.
The comparison between \selfcircle{B1} and \selfcircle{B3} showed that it was not helpful to distill two SSLs trained on different bona fide data.

We conclude that the large-scale vocoded data can be used for SSL continual training. The updated SSL can be used in a CM alone, and better performance is possible by using this SSL and its pre-trained version or distilling their differences into a student SSL.

\subsection{Experiment 2: large-scale vocoded data for CM fine-tuning}
\label{sec:exp2}
The second experiment investigates the effect of using large-scale vocoded data for downstream CM fine-tuning. 
Two versions of \selfcircle{P3} and \selfcircle{B1} were fine-tuned using the conventional ASVspoof 2019 LA training set \setTLAshort\  and the VoxCeleb2 dataset \setTVOXVOC\ with the corresponding bona fide data. 
Since it was infeasible to train the CM on the full set of \setTVOXVOC, we randomly sampled the same number of mini-batches as \setTLAshort\ in each epoch.

The results are listed in the right part of Table~\ref{tab_eer}. The comparison between \selfcircle{P3} and \selfcircle{P3-c} shows that using the large \setTVOXVOC\ for CM fine-tuning degraded the pooled EER. \selfcircle{P3-c}'s EER on \setEWFE\ dramatically increased to around 20\%, but the EER on \setELAHID\ was significantly reduced from 14.97\% to 9.53\%. A detailed analysis found that the improvement on \setELAHID\ came from the test data with GSM and PSTN codecs, on which the EER halved. This may be because \setTVOXVOC\ contains data from similar codecs.
However, fine-tuning the CM using \setTVOXVOC\ does not guarantee better performance on the test sets with no codec effect. 
Further investigation is needed to determine the reasons.

\subsection{Comparing results with existing studies}
\label{sec:exp3}
Table~\ref{tab:eer_comp} compares the results of \selfcircle{P3} with those reported in other studies on each test set.  
On the test sets from ASVspoof challenges, only CMs that have results on \emph{both} the original and non-speech-trimmed test sets are listed in the Table. As suggested by \cite{muller21_asvspoof} and \cite{LiuASVspoof2021}, CMs performing well on the standard ASVspoof challenges test sets need to be evaluated using test sets with non-speech trimmed.
Note that the best CM in \cite{Wang2023a} is equivalent to \selfcircle{B1} and is comparable with  \selfcircle{P3} in terms of model size and CM fine-tuning data.
\selfcircle{P3} outperforms other CMs on all test sets except the ASVspoof 2021 DF and LA. 
Additional analysis shows that the codec and compression in the ASVspoof 2021 LA and DF test sets remain challenging for \selfcircle{P3}, especially the test data processed by GSM and PSTN codecs.

\begin{table}[t!]
\caption{Comparison of EERs (\%) with other literature. Bolded \textbf{EER} marks best value(s) with a statistically significant margin from others. \cite{muller21_asvspoof} used a different tool to trim non-speech in {\setELATRIM}.
}
\vspace{-6mm}
\begin{center}
\footnotesize
\setlength{\tabcolsep}{2pt}{
\begin{tabular}{lrr}
\toprule
 &  {\setELA} & {\setELATRIM} \\ 
 \midrule
\cite{muller21_asvspoof} & 7.35  & *35.32 \\
\cite{Wang2023a} &  {2.21} & \textbf{3.79} \\ 
\selfcircle{P3} &  \textbf{1.91} & \textbf{3.28} \\ 
 \bottomrule
\end{tabular}
\quad
\begin{tabular}{lrr}
\toprule
 & {\setELAII} & {\setELAHID} \\ 
 \midrule
 \cite{LiuASVspoof2021}& \textbf{1.32} & $>$22  \\
\cite{Wang2023a} & 17.90  &  \textbf{14.57} \\ 
\selfcircle{P3} &  {15.92} & \textbf{14.97} \\ 
 \bottomrule
\end{tabular}\quad
\begin{tabular}{lrr}
\\
\toprule
 & {\setEDF} & {\setEDFHID} \\ 
 \midrule
 \cite{LiuASVspoof2021}& 15.64 & $>$20  \\
\cite{Wang2023a} & \textbf{5.04}  &  \textbf{7.78} \\ 
\selfcircle{P3}  & \textbf{5.67}  &  {8.84} \\ 
 \bottomrule
\end{tabular}\quad
\begin{tabular}{lr}
\\
\toprule
 & {\setEWFE}\\ 
 \midrule
 \cite{Wang2023}& 8.29 \\
\cite{Wang2023a} & {2.50}  \\ 
\selfcircle{P3}  & \textbf{1.30}  \\ 
 \bottomrule
\end{tabular}\quad
\begin{tabular}{lr}
\\
\toprule
 & {\setEDFE}\\ 
 \midrule
 \cite{pianese2022deepfake}& 15.00 \\
\cite{Wang2023a} & {7.55}  \\ 
\selfcircle{P3}  & \textbf{6.10}  \\ 
 \bottomrule
\end{tabular}
}
\vspace{-5mm}
\end{center}
\label{tab:eer_comp}
\end{table}

\section{Conclusion}
\label{sec:con}

Existing studies have shown the effectiveness of small-scale vocoded data in fine-tuning spoofing CMs. 
This study took one step further and generated more than 9k hours of vocoded data from VoxCeleb2. 
The vocoded data was used to continuously train the SSL model before integrating it into the CM or to fine-tune the entire CM using supervised learning.  

While the experiments found that directly fine-tuning the CM using the large-scale vocoded data was unsatisfactory, 
the CM using the continually trained SSL as the front end achieved a lower overall EER than the CM using an SSL pre-trained on bona fide data. Another finding is that computing the differences between the outputs of the continually trained and the pre-trained SSLs led to sparse and more discriminative features. 
On the basis of that observation, this study proposed a distillation method to encourage the SSL to produce sparse features. The CM using the distilled SSL outperformed the previously best model on challenging test sets, such as In-the-Wild. 
In short, the large-scale vocoded data is helpful in continually training the SSL for the anti-spoofing task.
Although further investigation is needed to improve the results on test data with codecs, we hope that the current findings and the trained SSL models can be helpful to the community.

\vfill\pagebreak

\pretolerance=1000

\bibliographystyle{IEEEbib}
\bibliography{library}

\begin{thebibliography}{10}

\bibitem{Wu2015}
Zhizheng Wu, Nicholas Evans, Tomi Kinnunen, Junichi Yamagishi, Federico Alegre,
  and Haizhou Li,
\newblock ``{Spoofing and Countermeasures for Speaker Verification: A
  survey},''
\newblock {\em Speech Communication}, vol. 66, pp. 130--153, feb 2015.

\bibitem{asvspoof2019_database}
Xin Wang, Junichi Yamagishi, Massimiliano Todisco, and Others,
\newblock ``{ASVspoof 2019: A Large-scale Public Database of Synthesized,
  Converted and Replayed Speech},''
\newblock {\em Computer Speech \& Language}, vol. 64, pp. 101114, nov 2020.

\bibitem{Wang2023a}
Xin Wang and Junichi Yamagishi,
\newblock ``{Spoofed Training Data for Speech Spoofing Countermeasure Can Be
  Efficiently Created Using Neural Vocoders},''
\newblock in {\em Proc. ICASSP}, 2023, pp. 1--5.

\bibitem{Sun2023CVPR}
Chengzhe Sun, Shan Jia, Shuwei Hou, and Siwei Lyu,
\newblock ``{AI-Synthesized Voice Detection Using Neural Vocoder Artifacts},''
\newblock in {\em Proc. CVPR Workshops}, 2023, pp. 904--912.

\bibitem{sanchez2014cross}
Jon Sanchez, Ibon Saratxaga, Inma Hernaez, Eva Navas, and Daniel Erro,
\newblock ``{A Cross-vocoder Study of Speaker Independent Synthetic Speech
  Detection Using Phase Information},''
\newblock in {\em Proc. Interspeech}, 2014.

\bibitem{sizov2015joint}
Aleksandr Sizov, Elie Khoury, Tomi Kinnunen, Zhizheng Wu, and S{\'{e}}bastien
  Marcel,
\newblock ``{Joint Speaker Verification and Antispoofing in the i-vector
  Space},''
\newblock {\em IEEE Transactions on Information Forensics and Security}, vol.
  10, no. 4, pp. 821--832, 2015.

\bibitem{chung18b_interspeech}
Joon~Son Chung, Arsha Nagrani, and Andrew Zisserman,
\newblock ``{VoxCeleb2}: Deep speaker recognition,''
\newblock in {\em Proc. Interspeech}, 2018, pp. 1086--1090.

\bibitem{mun2023towards}
Sung~Hwan Mun, Hye-jin Shim, Hemlata Tak, Xin Wang, Xuechen Liu, Md~Sahidullah,
  Myeonghun Jeong, Min~Hyun Han, Massimiliano Todisco, Kong~Aik Lee, and
  Others,
\newblock ``{Towards Single Integrated Spoofing-aware Speaker Verification
  Embeddings},''
\newblock {\em Proc. Interspeech}, pp. 3989--3993, 2023.

\bibitem{zeng2023improving}
Chang Zeng, Xin Wang, Xiaoxiao Miao, Erica Cooper, and Junichi Yamagishi,
\newblock ``{Improving Generalization Ability of Countermeasures for New
  Mismatch Scenario by Combining Multiple Advanced Regularization Terms},''
\newblock {\em Proc. Interspeech}, pp. 1998--2002, 2023.

\bibitem{NEURIPS2020_c5d73680}
Jungil Kong, Jaehyeon Kim, and Jaekyoung Bae,
\newblock ``{HiFi-GAN: Generative Adversarial Networks for Efficient and High
  Fidelity Speech Synthesis},''
\newblock in {\em Proc. NIPS}, 2020, vol.~33, pp. 17022--17033.

\bibitem{prenger2018waveglow}
Ryan Prenger, Rafael Valle, and Bryan Catanzaro,
\newblock ``{WaveGlow: A Flow-based Generative Network for Speech Synthesis},''
\newblock in {\em Proc. ICASSP}, 2019, pp. 3617--3621.

\bibitem{wangNSFall}
Xin Wang, Shinji Takaki, and Junichi Yamagishi,
\newblock ``{Neural Source-Filter Waveform Models for Statistical Parametric
  Speech Synthesis},''
\newblock {\em IEEE/ACM Transactions on Audio, Speech, and Language
  Processing}, vol. 28, pp. 402--415, 2020.

\bibitem{Mohamed2022}
Abdel-Rahman Mohamed, Hung-yi Lee, Lasse Borgholt, Jakob~D Havtorn, Joakim
  Edin, Christian Igel, Katrin Kirchhoff, Shang-Wen Li, Karen Livescu, Lars
  Maaloe, Tara~N Sainath, and Shinji Watanabe,
\newblock ``{Self-Supervised Speech Representation Learning: A Review},''
\newblock {\em IEEE Journal of Selected Topics in Signal Processing}, vol. 16,
  no. 6, pp. 1179--1210, oct 2022.

\bibitem{NEURIPS2020_92d1e1eb}
Alexei Baevski, Yuhao Zhou, Abdelrahman Mohamed, and Michael Auli,
\newblock ``{Wav2vec 2.0: A Framework for Self-Supervised Learning of Speech
  Representations},''
\newblock in {\em Proc. NIPS}, 2020, vol.~33, pp. 12449--12460.

\bibitem{tak2022automatic}
Hemlata Tak, Massimiliano Todisco, Xin Wang, Jee-weon Jung, Junichi Yamagishi,
  and Nicholas Evans,
\newblock ``{Automatic speaker verification spoofing and deepfake detection
  using wav2vec 2.0 and data augmentation},''
\newblock in {\em Proc. Odyssey}, 2022, pp. 112--119.

\bibitem{wang2021investigating}
Xin Wang and Junichi Yamagishi,
\newblock ``{Investigating Self-Supervised Front Ends for Speech Spoofing
  Countermeasures},''
\newblock in {\em Proc. Odyssey}, 2022, pp. 100--106.

\bibitem{martin2022vicomtech}
Juan~M Martin-Donas and Aitor Alvarez,
\newblock ``{The Vicomtech Audio Deepfake Detection System Based on Wav2vec2
  for the 2022 ADD Challenge},''
\newblock in {\em Proc. ICASSP}. IEEE, 2022, pp. 9241--9245.

\bibitem{LiuASVspoof2021}
Xuechen Liu, Xin Wang, Md~Sahidullah, Jose Patino, H{\'{e}}ctor Delgado, Tomi
  Kinnunen, Massimiliano Todisco, Junichi Yamagishi, Nicholas Evans, Andreas
  Nautsch, and Kong~Aik Lee,
\newblock ``{ASVspoof 2021: Towards Spoofed and Deepfake Speech Detection in
  the Wild},''
\newblock {\em IEEE/ACM Transactions on Audio, Speech, and Language
  Processing}, vol. 31, pp. 2507--2522, 2023.

\bibitem{muller21_asvspoof}
Nicolas M{\"{u}}ller, Franziska Dieckmann, Pavel Czempin, Roman Canals,
  Konstantin B{\"{o}}ttinger, and Jennifer Williams,
\newblock ``{Speech is Silver, Silence is Golden: What do ASVspoof-trained
  Models Really Learn?},''
\newblock in {\em Proc. ASVspoof Challenge workshop}, 2021, pp. 55--60.

\bibitem{zhang21da_interspeech}
Yuxiang Zhang, Wenchao Wang, and Pengyuan Zhang,
\newblock ``{The Effect of Silence and Dual-Band Fusion in Anti-Spoofing
  System},''
\newblock in {\em Proc. Interspeech}, 2021, pp. 4279--4283.

\bibitem{frank2021wavefake}
Joel Frank and Lea Sch{\"{o}}nherr,
\newblock ``{WaveFake: A Data Set to Facilitate Audio DeepFake Detection},''
\newblock in {\em Proc. NeurIPS Datasets and Benchmarks 2021}, 2021.

\bibitem{muller2022does}
Nicolas~M M{\"{u}}ller, Pavel Czempin, Franziska Dieckmann, Adam Froghyar, and
  Konstantin B{\"{o}}ttinger,
\newblock ``{Does Audio Deepfake Detection Generalize?},''
\newblock {\em Proc. Interspeech}, pp. 2783--2787, 2022.

\bibitem{Tak2021}
Hemlata Tak, Madhu~R Kamble, Jose Patino, Massimiliano Todisco, and Nicholas
  W~D Evans,
\newblock ``{RawBoost: A Raw Data Boosting and Augmentation Method applied to
  Automatic Speaker Verification Anti-Spoofing},''
\newblock in {\em Proc. ICASSP}, 2022, pp. 6382--6386.

\bibitem{kingma2014adam}
Diederik~P Kingma and Jimmy Ba,
\newblock ``{Adam: A method for stochastic optimization},''
\newblock in {\em Proc. ICLR}, 2014.

\bibitem{conneau21_interspeech}
Alexis Conneau, Alexei Baevski, Ronan Collobert, Abdelrahman Mohamed, and
  Michael Auli,
\newblock ``{Unsupervised Cross-Lingual Representation Learning for Speech
  Recognition},''
\newblock in {\em Proc. Interspeech}, 2021, pp. 2426--2430.

\bibitem{ott2019fairseq}
Myle Ott, Sergey Edunov, Alexei Baevski, Angela Fan, Sam Gross, Nathan Ng,
  David Grangier, and Michael Auli,
\newblock ``{FairSeq}: A fast, extensible toolkit for sequence modeling,''
\newblock in {\em Proceedings of NAACL-HLT 2019: Demonstrations}, 2019.

\bibitem{Wang2023}
Xin Wang and Junichi Yamagishi,
\newblock ``Investigating active-learning-based training data selection for
  speech spoofing countermeasure,''
\newblock in {\em Proc. SLT}, 2023, pp. 585--592.

\bibitem{pianese2022deepfake}
Alessandro Pianese, Davide Cozzolino, Giovanni Poggi, and Luisa Verdoliva,
\newblock ``{Deepfake audio detection by speaker verification},''
\newblock {\em arXiv preprint arXiv:2209.14098}, 2022.

\end{thebibliography}

\newpage
\clearpage
\appendix
\section{Appendix}

\begin{table}[h!]
\caption{Test set EERs (\%) of ablation study. Teacher SSLs are \sslxlsr\ and \sslvox. Dataset used to fine-tune CM is \setTLAVOC.}
\vspace{-6mm}
\begin{center}
\footnotesize
\setlength{\tabcolsep}{4pt}
{
\begin{tabular}{rr||rrrrr}
\toprule
\multirow{5}{*}{\rotatebox{90}{CM}} & ID & \selfcircle{P3} & \selfcircle{P3-2} & \selfcircle{P3-3} & \selfcircle{P3-4} & \selfcircle{P3-5}  \\
\cmidrule{2-7}
  & Student SSL init.    & \multicolumn{4}{c}{\sslxlsr} & random \\ 
\cmidrule(lr){2-6}\cmidrule(lr){7-7}
& $\lambda_{\text{dis}}$ & 100 & 10 & 1000 & \multicolumn{2}{c}{100} \\
\cmidrule(lr){2-3}\cmidrule(lr){4-4}\cmidrule(lr){5-5}\cmidrule(lr){6-7}
& Distil. hidden & $\times$ & $\times$ & $\times$ & $\checkmark$ & $\times$ \\
\midrule
\multirow{11}{*}{\rotatebox{90}{Test sets}}          &   \setELA   & \cellcolor[rgb]{0.98, 0.98, 0.98} 1.91 & \cellcolor[rgb]{0.99, 0.99, 0.99} 1.65 & \cellcolor[rgb]{0.93, 0.93, 0.93} 6.99 & \cellcolor[rgb]{0.94, 0.94, 0.94} 6.38 & \cellcolor[rgb]{0.59, 0.59, 0.59} 30.22\\ 
  &  \setELAII  & \cellcolor[rgb]{0.79, 0.79, 0.79} 15.92 & \cellcolor[rgb]{0.79, 0.79, 0.79} 16.21 & \cellcolor[rgb]{0.59, 0.59, 0.59} 27.18 & \cellcolor[rgb]{0.59, 0.59, 0.59} 30.89 & \cellcolor[rgb]{0.59, 0.59, 0.59} 33.88\\ 
  &   \setEDF   & \cellcolor[rgb]{0.95, 0.95, 0.95} 5.67 & \cellcolor[rgb]{0.87, 0.87, 0.87} 11.27 & \cellcolor[rgb]{0.75, 0.75, 0.75} 18.08 & \cellcolor[rgb]{0.67, 0.67, 0.67} 21.58 & \cellcolor[rgb]{0.59, 0.59, 0.59} 38.03\\ 
  & \setELATRIM & \cellcolor[rgb]{0.97, 0.97, 0.97} 3.28 & \cellcolor[rgb]{0.97, 0.97, 0.97} 3.10 & \cellcolor[rgb]{0.90, 0.90, 0.90} 9.56 & \cellcolor[rgb]{0.90, 0.90, 0.90} 9.48 & \cellcolor[rgb]{0.59, 0.59, 0.59} 40.21\\ 
  & \setELAHID  & \cellcolor[rgb]{0.81, 0.81, 0.81} 14.97 & \cellcolor[rgb]{0.76, 0.76, 0.76} 17.40 & \cellcolor[rgb]{0.64, 0.64, 0.64} 22.96 & \cellcolor[rgb]{0.59, 0.59, 0.59} 25.10 & \cellcolor[rgb]{0.59, 0.59, 0.59} 40.87\\ 
  & \setEDFHID  & \cellcolor[rgb]{0.90, 0.90, 0.90} 8.84 & \cellcolor[rgb]{0.84, 0.84, 0.84} 13.14 & \cellcolor[rgb]{0.79, 0.79, 0.79} 15.91 & \cellcolor[rgb]{0.70, 0.70, 0.70} 20.61 & \cellcolor[rgb]{0.59, 0.59, 0.59} 39.75\\ 
  &  \setEWFE   & \cellcolor[rgb]{0.99, 0.99, 0.99} 1.30 & \cellcolor[rgb]{0.99, 0.99, 0.99} 1.52 & \cellcolor[rgb]{0.95, 0.95, 0.95} 5.69 & \cellcolor[rgb]{0.59, 0.59, 0.59} 27.49 & \cellcolor[rgb]{0.59, 0.59, 0.59} 55.94\\ 
  &  \setEDFE   & \cellcolor[rgb]{0.94, 0.94, 0.94} 6.10 & \cellcolor[rgb]{0.88, 0.88, 0.88} 10.87 & \cellcolor[rgb]{0.80, 0.80, 0.80} 15.60 & \cellcolor[rgb]{0.68, 0.68, 0.68} 21.45 & \cellcolor[rgb]{0.59, 0.59, 0.59} 39.97\\ 
\cmidrule{2-7}  
  &  \setGAll   & \cellcolor[rgb]{0.89, 0.89, 0.89} 9.98 & \cellcolor[rgb]{0.87, 0.87, 0.87} 11.14 & \cellcolor[rgb]{0.71, 0.71, 0.71} 20.09 & \cellcolor[rgb]{0.68, 0.68, 0.68} 21.39 & \cellcolor[rgb]{0.59, 0.59, 0.59} 37.42\\ 
\bottomrule
\end{tabular}
}
\label{tab_eer_2}
\end{center}
\end{table}

\subsection{Ablation study on distilling configuration}

An ablation study was conducted using the CMs listed below. They are the same as \selfcircle{P3} except that
\begin{itemize}
\item \selfcircle{P3-2} used $\lambda_{\text{dis}} = 10$,
\item \selfcircle{P3-3} used $\lambda_{\text{dis}} = 1000$,
\item \selfcircle{P3-4} distilled the differences between hidden features of teacher SSLs rather than SSLs' outputs and
\item \selfcircle{P3-5} randomly initialized the student SSL before distilling.
\end{itemize}
For \selfcircle{P3-4}, the hidden features refer to the output from the 24 transformer blocks in the SSL front end. In contrast, the SSL's output refers to the final output from the layer normalization following the last transformer block.

Results in Table~\ref{tab_eer_2} show that $\lambda_{\text{dis}}=100$ led to better performance than $\lambda_{\text{dis}}=10$ or $1000$. When using $\lambda_{\text{dis}}=100$, the amplitude of the error $\lambda_{\text{dis}}\mathcal{L}_{\text{dis}}$ is comparable to that of $\lambda_{\text{CF}}\mathcal{L}_{\text{CF}}$.  Distilling hidden features or random initializing the student SSL degraded the performance. Hence, this experiment found that it is better to initialize the student SSL using a pre-trained set of parameters and apply distillation on the SSL's output layer.

\begin{table*}[t!]
\caption{Test set EERs (\%). Darker cell color indicates a higher EER value. 
\selfcircle{B1-b}, \selfcircle{B1-c}, \selfcircle{B1-d} are the same as \selfcircle{B1} except that they were fine-tuned on \setTLAshort\, \setTVOXVOC, and \setTLAVOC+\setTVOXVOC, respectively. Similarly, \selfcircle{P3-b}, \selfcircle{P3-c}, \selfcircle{P3-d} are variants of \selfcircle{P3}. Results apart from those of \selfcircle{B1-d} and \selfcircle{P3-d} are copied from Table~\ref{tab_eer}. 
}
\vspace{-6mm}
\begin{center}
\footnotesize
\setlength{\tabcolsep}{8pt}
{
\begin{tabular}{rr||rr|rr|rr|rr}
\toprule
 \multirow{3}{*}{\rotatebox{90}{CM}}       &          ID        & \selfcircle{B1}   & \selfcircle{P3}        & \selfcircle{B1-b}   & \selfcircle{P3-b}          & \selfcircle{B1-c}  &  \selfcircle{P3-c} & \selfcircle{B1-d}  &  \selfcircle{P3-d} \\ 
\cmidrule{2-10}
& {Front end SSL(s)} & 
\textcolor{white}{-\ssllv}\sslxlsr & \sslxlsr, \sslvox & \textcolor{white}{-\ssllv}\sslxlsr & \sslxlsr, \sslvox  & \textcolor{white}{-\ssllv}\sslxlsr & \sslxlsr, \sslvox  & \textcolor{white}{-\ssllv}\sslxlsr & \sslxlsr, \sslvox  \\
 &  
SSL distilling & -  & $\checkmark$ & -  & $\checkmark$ & -  & $\checkmark$ & -  & $\checkmark$\\
\midrule
\multicolumn{2}{r||}{Data for fine-tune CM}  &  \multicolumn{2}{c|}{ \setTLAVOC}  &  \multicolumn{2}{c|}{ \setTLAshort}  &   \multicolumn{2}{c|}{ \setTVOXVOC}   &  \multicolumn{2}{c}{ \setTLAVOC\setTVOXVOC} \\
\midrule
\midrule
\multirow{11}{*}{\rotatebox{90}{Test sets}}     
  &   \setELA   & \cellcolor[rgb]{0.96, 0.96, 0.96} 3.45 & \cellcolor[rgb]{0.98, 0.98, 0.98} 1.91 & \cellcolor[rgb]{1.00, 1.00, 1.00} 0.22 & \cellcolor[rgb]{1.00, 1.00, 1.00} 0.13 & \cellcolor[rgb]{0.96, 0.96, 0.96} 3.59 & \cellcolor[rgb]{0.96, 0.96, 0.96} 3.71 & \cellcolor[rgb]{0.89, 0.89, 0.89} 7.72 & \cellcolor[rgb]{0.64, 0.64, 0.64} 18.15\\ 
  &  \setELAII  & \cellcolor[rgb]{0.66, 0.66, 0.66} 17.59 & \cellcolor[rgb]{0.72, 0.72, 0.72} 15.92 & \cellcolor[rgb]{0.97, 0.97, 0.97} 2.69 & \cellcolor[rgb]{0.96, 0.96, 0.96} 3.29 & \cellcolor[rgb]{0.73, 0.73, 0.73} 15.22 & \cellcolor[rgb]{0.80, 0.80, 0.80} 12.37 & \cellcolor[rgb]{0.81, 0.81, 0.81} 12.09 & \cellcolor[rgb]{0.59, 0.59, 0.59} 24.78\\ 
  &   \setEDF   & \cellcolor[rgb]{0.92, 0.92, 0.92} 6.53 & \cellcolor[rgb]{0.93, 0.93, 0.93} 5.67 & \cellcolor[rgb]{0.95, 0.95, 0.95} 4.27 & \cellcolor[rgb]{0.96, 0.96, 0.96} 3.45 & \cellcolor[rgb]{0.92, 0.92, 0.92} 5.99 & \cellcolor[rgb]{0.96, 0.96, 0.96} 3.31 & \cellcolor[rgb]{0.93, 0.93, 0.93} 5.61 & \cellcolor[rgb]{0.75, 0.75, 0.75} 14.52\\ 
    \cmidrule{2-10}
  & \setELATRIM & \cellcolor[rgb]{0.97, 0.97, 0.97} 2.69 & \cellcolor[rgb]{0.96, 0.96, 0.96} 3.28 & \cellcolor[rgb]{0.90, 0.90, 0.90} 7.37 & \cellcolor[rgb]{0.90, 0.90, 0.90} 7.37 & \cellcolor[rgb]{0.97, 0.97, 0.97} 2.74 & \cellcolor[rgb]{0.96, 0.96, 0.96} 3.63 & \cellcolor[rgb]{0.96, 0.96, 0.96} 3.36 & \cellcolor[rgb]{0.83, 0.83, 0.83} 10.85\\ 
  & \setELAHID  & \cellcolor[rgb]{0.76, 0.76, 0.76} 13.93 & \cellcolor[rgb]{0.74, 0.74, 0.74} 14.97 & \cellcolor[rgb]{0.72, 0.72, 0.72} 15.56 & \cellcolor[rgb]{0.59, 0.59, 0.59} 24.23 & \cellcolor[rgb]{0.85, 0.85, 0.85} 10.14 & \cellcolor[rgb]{0.86, 0.86, 0.86} 9.53 & \cellcolor[rgb]{0.90, 0.90, 0.90} 7.62 & \cellcolor[rgb]{0.75, 0.75, 0.75} 14.61\\ 
  & \setEDFHID  & \cellcolor[rgb]{0.87, 0.87, 0.87} 8.89 & \cellcolor[rgb]{0.87, 0.87, 0.87} 8.84 & \cellcolor[rgb]{0.87, 0.87, 0.87} 9.16 & \cellcolor[rgb]{0.76, 0.76, 0.76} 13.95 & \cellcolor[rgb]{0.87, 0.87, 0.87} 9.03 & \cellcolor[rgb]{0.89, 0.89, 0.89} 7.77 & \cellcolor[rgb]{0.90, 0.90, 0.90} 7.14 & \cellcolor[rgb]{0.78, 0.78, 0.78} 13.42\\ 
  &  \setEWFE   & \cellcolor[rgb]{0.90, 0.90, 0.90} 7.33 & \cellcolor[rgb]{0.99, 0.99, 0.99} 1.30 & \cellcolor[rgb]{0.59, 0.59, 0.59} 23.75 & \cellcolor[rgb]{0.73, 0.73, 0.73} 15.44 & \cellcolor[rgb]{0.78, 0.78, 0.78} 13.41 & \cellcolor[rgb]{0.59, 0.59, 0.59} 24.17 & \cellcolor[rgb]{0.88, 0.88, 0.88} 8.31 & \cellcolor[rgb]{0.59, 0.59, 0.59} 27.75\\ 
  &  \setEDFE   & \cellcolor[rgb]{0.91, 0.91, 0.91} 6.78 & \cellcolor[rgb]{0.92, 0.92, 0.92} 6.10 & \cellcolor[rgb]{0.77, 0.77, 0.77} 13.52 & \cellcolor[rgb]{0.80, 0.80, 0.80} 12.32 & \cellcolor[rgb]{0.91, 0.91, 0.91} 6.90 & \cellcolor[rgb]{0.91, 0.91, 0.91} 7.00 & \cellcolor[rgb]{0.88, 0.88, 0.88} 8.36 & \cellcolor[rgb]{0.59, 0.59, 0.59} 20.36\\ 
  \cmidrule{2-10}
  &  \setGAll   & \cellcolor[rgb]{0.83, 0.83, 0.83} 11.13 & \cellcolor[rgb]{0.85, 0.85, 0.85} 9.98 & \cellcolor[rgb]{0.79, 0.79, 0.79} 12.76 & \cellcolor[rgb]{0.80, 0.80, 0.80} 12.50 & \cellcolor[rgb]{0.83, 0.83, 0.83} 10.92 & \cellcolor[rgb]{0.80, 0.80, 0.80} 12.26 & \cellcolor[rgb]{0.83, 0.83, 0.83} 11.00 & \cellcolor[rgb]{0.59, 0.59, 0.59} 20.90\\   \bottomrule
\end{tabular}
}
\label{tab_eer_database_app}
\vspace{-6mm}
\end{center}
\end{table*}

\begin{figure}[t!]
\begin{center}
\includegraphics[trim=0 10 0 0, width=\columnwidth]{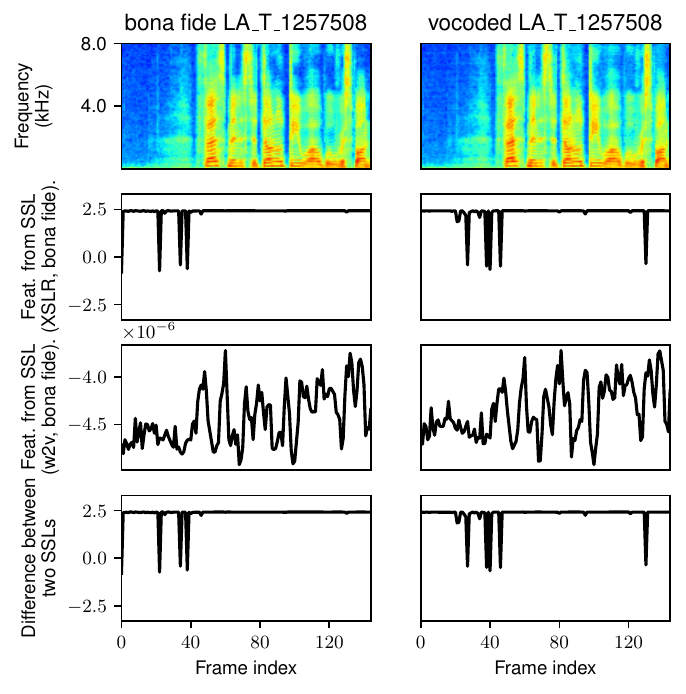}
\caption{Spectrograms of bona fide and vocoded utterances (top row), output features from \sslxlsr\  (2nd row) and \ssllv\  (3rd row), and their differences (bottom row). }
\label{fig_twossl_base}
\vspace{-5mm}
\end{center}
\end{figure}

\begin{figure}[t!]
\begin{center}
\includegraphics[trim=0 10 0 0, width=\columnwidth]{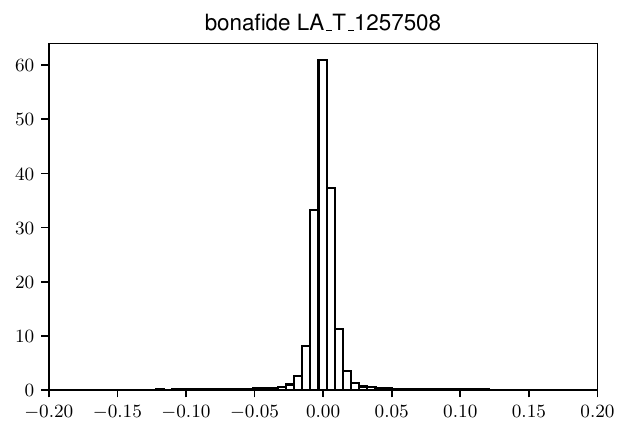}
\includegraphics[trim=0 10 0 0, width=\columnwidth]{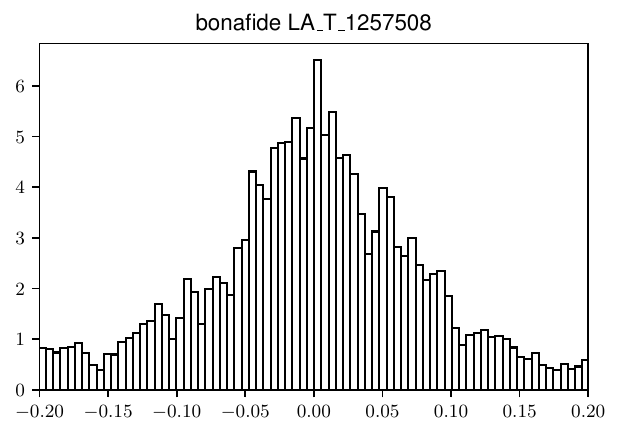}
\caption{Histogram of differential features between \sslxlsr\  and \sslvox\ (top) and that between \sslxlsr\  and \ssllv\  (bottom).}
\label{fig_twossl_hist}
\vspace{-5mm}
\end{center}
\end{figure}

\subsection{Features from different SSL models}
In Section~\ref{sec:exp1}, we mentioned that the outputs from \sslxlsr\ and \ssllv\ used by \selfcircle{B2} were quite different. Figure~\ref{fig_twossl_base} plots the features from the two SSLs, using the same style as Figure~\ref{fig_twossl}. Only a single dimension of the feature is plotted, and this dimension is the one with the largest variance in the differential features between the two SSLs' outputs, which happened to be the same dimension as that in Figure~\ref{fig_twossl}.  

It is observed that \ssllv's output has a much smaller dynamic range, and the difference between \sslxlsr\ and \ssllv\ is similar to the output from \sslxlsr. Notice that the differential feature is centered around 2.5 rather than 0.0. 

Figure~\ref{fig_twossl_hist} shows the histogram of the differential features pooled over all the 1,024 dimensions and time steps on a bona fide utterance. The differential features between  \sslxlsr\  and \sslvox\ are closer to 0 than those from \sslxlsr\  and \ssllv.

\subsection{Additional experiments on databases for CM finetuning}

Section~\ref{sec:exp2} compared the CMs finetuned on databases \setTLAshort\  and  \setTVOXVOC\. We also tried to combine the databases (though not all the possible combinations) and put the results in Table~\ref{tab_eer_database_app}. The new results are EERs of \selfcircle{B1-d} and \selfcircle{P3-d}.
Combining two databases for fine-tuning does not necessarily improve the performance. The reason for the poor performance of \selfcircle{P3-d} is currently unknown.


\end{document}